\documentclass[final,5p,times,twocolumn,authoryear]{elsarticle}

\usepackage{amssymb}
\usepackage{lipsum}

\newcommand{\mnras}{Mon. Not. R. Astron. Soc.}
\newcommand{\apjl}{Astrophys. J. Lett.}
\newcommand{\apj}{Astrophys. J.}

\newcommand{\aap}{Astron. \& Astrophys.}

\newcommand{\nar}{New Astronomy Reviews}

\newcommand{\nat}{Nature}
\newcommand{\na}{Nature Astronomy}

\journal{High Energy Astrophysics}

\begin{document}

\begin{frontmatter}



\title{\bf On the Duration of Gamma-Ray Bursts}


\author[1,2]{Bing Zhang}
\ead{bing.zhang@unlv.edu}
\affiliation[1]{organization={The Nevada Center for Astrophysics, University of Nevada, Las Vegas},
            city={Las Vega },
            postcode={89154}, 
            state={NV},
            country={USA}}
\affiliation[2]{organization={Department of Physics and Astronomy, University of Nevada, Las Vegas},
            city={Las Vega },
            postcode={89154}, 
            state={NV},
            country={USA}}

\begin{abstract}
Recently, a short-duration GRB with supernova association (GRB 200826A) and two long-duration GRBs with kilonova associations (GRB 211211A and GRB 230307A) have been detected, which demolished the hope for a tidy connection between GRB duration and their progenitor systems. Here I summarize various physical factors that can shape the duration of a GRB and propose that the duration of a GRB can be defined by four factors: progentor, central engine, emitter, and geometry. The progenitor-defined duration is only relevant when the central engine is powered by accretion and when the modifications by other factors are not important. The untidy situation of duration - progenitor mismatches suggests that other factors likely play important roles in defining GRB duration at least in some GRBs. In particular, a GRB  may not be powered by accretion but rather by a millisecond magnetar at least for some GRBs. The complicated lightcurve of GRB 211211A suggests both progenitor- and engine-defined durations, which may require a new type of progenitor system involving a white dwarf - neutron star merger with a magnetar merger product. The single broad pulse lightcurve with well-behaved energy-dependent behavior of GRB 230307A suggests an emitter-defined long duration. The central engine timescale may be short enough to be accommodated within the framework of a standard binary neutron star merger. Its spiky lightcurve with fast variability as well as extended X-ray emission suggest the existence of mini-jets in the global dissipation region, powered by an underlying magnetar.
\end{abstract}




\end{frontmatter}




\section{Introduction}
\label{sec:intro}

The duration of gamma-ray bursts (GRBs), typically measured by $T_{90}$ (the duration during which the fluence of the GRB increases from 5\% to 95\%), shows a bimodal distribution with a dip around 2-4 seconds depending on the detectors' energy band and sensitivity \citep{kouveliotou93,qin13,vonKienlin20}. This separates GRBs to two phenomenological categories: short- ($T_{90} < 2$ s) and long- ($T_{90} > 2$ s) duration GRBs.

On the other hand, multi-wavelength, multi-messenger observations suggest that there are two physically distinct categories \citep{zhangnature06,zhang09}:
\begin{itemize}
\item Compact star GRBs (Type I): those related to binary neutron star mergers \citep{paczynski86,eichler89,paczynski91,narayan92}, typically short, often associated with gravitational waves or kilonovae, but not associated with active star formation;
\item Massive star GRBs (Type II): those related to massive star core collapse \citep{woosley93,paczynski98,macfadyen99}, typically long, often associated with broad-line Type Ic supernovae and active star formation. 
\end{itemize}

The cozy one-to-one correspondence for ``short = Type I'' and ``long = Type II'' has been progressively muddled over the years. The situation is worsen recently with the discoveries of three peculiar GRBs, which completely demonished the hope of a clean separation between the two categories. 
\begin{itemize}
\item The short-duration, Type-II GRB 200826A had a 1-second-duration sharp pulse but with a supernova association \citep{ahumada21,zhangbb21}.
\item The long-duration, Type-I GRB 211211A had a total duration of $\sim 68$ s, with a $\sim 13$-s hard episode and a $\sim 55$-s extended emission \citep{yangj22}, but was associated with a kilonova \citep{rastinejad22}. 
\item Another long-duration, Type-I GRB 230307A had a duration of $\sim 42$ s \citep{sun23} with a global energy-dependent, self-similar single-pulse profile \citep{yi23} with superposed rapid variability,  but was also associated with a kilonova \citep{levan24,yang24}. 
\end{itemize}

These observations beg the question regarding how much one should trust the usage of duration as the main criterion to infer the progenitor system of a GRB. The purpose of this Letter is to clarify various physical factors that can define/modify the observed GRB duration. In Section \ref{sec:factors}, I discuss GRB duration defined by progenitor (Sect. \ref{sec:progenitor}), central engine (Sect. \ref{sec:engine}), emitter (Sect. \ref{sec:emitter}), and geometry (Sect. \ref{sec:geometry}). In Section \ref{sec:case}, I present a case study of the three peculiar GRBs and discuss the likely physical scenarios to interpret these events. Our results are summarized in Section \ref{sec:summary}. 

\section{GRB duration defined/modified by various physical factors}\label{sec:factors}

\subsection{Progenitor-defined duration}\label{sec:progenitor}

The rationale behind duration -- progenitor correspondence lies in the assumption that GRB jets are accretion powered, with the accretion fallback timescale related to the density of the progenitor star \citep[e.g.][]{zhang18}. 
Most generally, the GRB duration for an accretion-powered engine may be defined as
\begin{equation}
T_{\rm GRB} \simeq {\rm max} (t_{\rm ff}, t_{\rm acc}) - t_{\rm bo},
\label{eq:TGRB}
\end{equation}
where $t_{\rm ff}$ is the free fall timescale of the progenitor star, which defines how long it takes for the available material falls onto the disk (e.g. at the outer boundary), $t_{\rm acc}$ is the characteristic accretion timescale for a mass element traveling from the outer boundary of the disk to the central object, and $t_{\rm bo}$ is the timescale for the jet to breakout the envelope of the progenitor system. The central object is usually assumed as a stellar-mass black hole \citep{woosley93,macfadyen99,popham99,meszaros99,narayan01,lei13,lei17,liu17}, even though a millisecond magnetar could also serve as an accretor \citep{zhangdai08,zhangdai09,zhangdai10,metzger18b} before its other energy powers kick in.

The free-fall timescale of the progenitor system can be estimated as 
\begin{eqnarray}
    t_{\rm ff} & \simeq & \left(\frac{3\pi}{32 G \bar\rho}\right)^{1/2} \nonumber \\
    &\simeq & (180 \ {\rm s}) \ \left(\frac{\bar\rho}{100~{\rm g~cm^{-3}}}\right)^{-1/2} \nonumber \\
    &\simeq & (0.02 \ {\rm s}) \ \left(\frac{\bar\rho}{10^{10}~{\rm g~cm^{-3}}}\right)^{-1/2},\end{eqnarray}
where the characteristic values of $\bar\rho$ for a massive star and a binary neutron star merger ejecta are used. One can see that the predicted characteristic values of $t_{\rm ff}$ well match the two progenitor types. This was the main motivation for the well-known duration - progenitor connection. 

The accretion timescale may be estimated as the viscous timescale of the disk, which is the fallback timescale of the {\em disk} (not progenitor star) stretched by the viscosity parameter $\alpha<1$ by a factor of $\alpha^{-1}$. Adopting a typical density at a large radius, e.g. $\sim 100 r_s$ ($r_s$ is the Schwarzschild radius of the central accreting black hole), one can get a loose upper limit of the accretion timescale
\begin{eqnarray}
    t_{\rm acc} & < & \alpha^{-1} \left(\frac{3\pi}{32 G \rho_d}\right)^{1/2} \nonumber \\
    &\simeq & (3 \ {\rm s})  \ \alpha_{-1}^{-1} \left(\frac{\rho_d}{5\times 10^7 \ {\rm g~cm^{-3}}}\right)^{-1/2},
\label{eq:tacc}
\end{eqnarray}
where the typical value $\rho_d (100 r_s) \simeq (5\times 10^7) \ {\rm g~cm^{-3}}$ has been adopted for the parameter set $M=3 M_\odot$ and $\dot M = 0.1 \ (M_\odot/{\rm s})$, which applies for both a neutrino-dominated-accretion flow and an advection-dominated-accretion flow \citep{narayan01,yuan12}.

More precise estimates of $t_{\rm acc}$ should take into account radius-dependent density and other parameters, as well as the physical regimes of the accretion disk, i.e. whether the disk is convection dominated or neutrino-cooling dominated, and whether the pressure is gas or radiation dominated \citep[e.g.][]{narayan01}. Without repeating the derivations, we list the results of \cite{narayan01} as follows:
\begin{equation}
    t_{\rm acc} \simeq \left\{
    \begin{array}{ll}
       (4.2 \times 10^{-4}\ {\rm s}) \ \alpha_{-1}^{-1} m_3 r_{\rm out}^{3/2},  & r_{\rm out} > r_1, \\
       (2.8 \times 10^{-2}\ {\rm s}) \ \alpha_{-1}^{-6/5} m_3^{6/5} r_{\rm out}^{4/5},  & r_2<r_{\rm out} < r_1, \\
    (2.8 \times 10^{-3}\ {\rm s}) \ \alpha_{-1}^{-1} m_3^{13/7} m_d^{-2/7} r_{\rm out}^{3/2},  & r_{\rm out} < r_2, 
    \end{array} \\
    \right. 
\end{equation}
where
\begin{eqnarray}
    r_1 & = & 118 \ \alpha_{-1}^{-2/7} m_3^{-1} m_d^{3/7}, \\
    r_2 & = & 26.2 \alpha_1^{-2/7} m_3^{-46/49} m_d^{20/49} 
\end{eqnarray}
separate various regimes, with $r_1$ separating CDAF (above $r_1$) and NDAF (below $r_1$) and $r_2$ separating gas pressure dominance (above $r_2$) and radiation pressure dominance (below $r_2$). Here parameters are normalized, with accretor mass $m_3 = M/(3 M_\odot)$, disk mass $m_d = M_d/M_\odot$, and $r_{\rm out} = R_{\rm out}/r_s$. One can see for typical parameters, $t_{\rm acc}$ is shorter than 1 s, suggesting that without significant fallback (like the case of compact binary mergers), the jet duration should be quite short, consistent with $T_{90} < 2$ s for a large fraction of Type I GRBs. 

Assuming that the accretion timescale corresponds to the jet operation timescale, the final observed GRB duration only accounts for the duration when the jet is successful. The jet breakout time can be estimated as 
\begin{eqnarray}
    t_{\rm bo} & \simeq & R_{\rm env} / v_{\rm jh} \nonumber \\
    &\simeq & (10 \ {\rm s})  \left(\frac{R_{\rm env}}{3\times 10^{10} \ {\rm cm}}\right) \left(\frac{v_{\rm jh}}{0.1 c}\right) \nonumber \\
    &\simeq & (1 \ {\rm s})  \left(\frac{R_{\rm env}}{3\times 10^{10} \ {\rm cm}}\right) \left(\frac{v_{\rm jh}}{c}\right), 
    \end{eqnarray}
where the two typical values are defined by the characteristic parameter values of the two types of progenitors: For a Type II GRB, the envelope radius $R_{\rm env}$ is defined by the radius of the Wolf-Rayet progenitor star. Due to the high density in the envelope, the jet head travels with a non-relativistic speed, and we normalize it to $0.1 c$. For a compact binary merger remnant, the size of the envelope is defined by the delay time between the merger and the launch of the jet, which is adopted as $\sim 1$ s\footnote{The delay time between the merger and jet launching in GW170817/GRB 170817A is not well measured. The observed delay is about 1.7 s \citep{GW170817/GRB170817A}. Some authors \citep[e.g.][]{gill19} argued for a delay of the order of $\sim 1$ s. However, since the observed delay time is comparable to the duration of the burst itself \citep{zhangbb18b}, it is possible that the observed delay time is mostly attributed to the time for the jet to propagate to the dissipation radius, so that the jet launching time delay could be much shorter than 1 s \citep{zhang19b}. }. Because the envelope density is low, the jet propagation speed is likely close to speed of light. 

Based on Equation (\ref{eq:TGRB}), one can get the following consistent picture for a population of ``well-behaved GRBs'':
\begin{itemize}
 \item For Type II GRBs, one has $t_{\rm ff} \gg t_{\rm acc}$, so the duration is mostly defined by $T_{\rm GRB} \simeq t_{\rm ff} - t_{\rm bo}$. Consider a range of progenitor parameters. It is likely that in most cases one has $t_{\rm ff} \gg t_{\rm bo}$ so that the GRB duration is long. Under certain conditions, $t_{\rm ff} \gtrsim t_{\rm bo}$ is satisfied, so a duration shorter than 10 s and even 2 s is allowed \citep{bromberg11b}. If  $t_{\rm ff} < t_{\rm bo}$ is satisfied, the jet becomes unsuccessful, which would power a low-luminosity GRB or X-ray flash with emission dominated by shock breakout \citep{meszarosrees01,campana06,nakar12}.
 \item For Type I GRBs from binary neutron star mergers, one typically has $t_{\rm ff} \ll t_{\rm acc}$, so the duration is mainly defined by $T_{\rm GRB} \simeq t_{\rm acc}-t_{\rm bo}$. For typical parameters, we get $T_{\rm GRB} < 2$ s if the GRB jets are powered by accretion. 
\end{itemize}

\subsection{Engine-defined duration}\label{sec:engine}

If the GRB central engine is not an accreting black hole, but a millisecond magnetar as widely discussed as an alternative engine for both types of GRBs \citep{usov92,thompson94,dailu98a,zhangmeszaros01a,metzger11}, the GRB duration would not sensitively depend on the progenitor type. The duration defined by the magnetar engine activity timescale is called ``engine-defined duration'' in the following\footnote{Strictly speaking, the duration defined in \S\ref{sec:progenitor} is also defined by the accretion timescale of the central engine. However, because that timescale is closely related to the progenitor type, I define it as ``progenitor-defined duration''.}.

The duration of a burst with a magnetar engine can be as long as the life time (for a supramassive neutron star) or the spindown time (for a stable neutron star) of the magnetar. The exact mechanism for a millisecond magnetar to produce GRB prompt emission is not clear. At least three mechanisms have been discussed in the literature:
\begin{itemize}
    \item Accretion: Within this scenario \citep{zhangdai08,zhangdai09,zhangdai10,metzger18b}, the duration is still essentially defined by the accretion timescale as discussed in Section \ref{sec:progenitor}, with possible modifications due to the magnetic barrier of the magnetar that may delay and prolong the total duration to some degree. 
    \item Magnetic bubble emission from a differentially rotating neutron star: Such a mechanism was proposed for both Type II GRBs to account for sporadic prompt emission \citep{kluzniak98,ruderman00} and Type I GRBs \citep{dailu98a,dai06} to account for prompt and late engine activities. The duration of the burst is defined by the timescale before the differential rotation is damped, which can be from seconds to tens of seconds \citep{shapiro00}. Therefore, for both Type II and Type I GRBs, the duration of the burst can be ``long''.
    \item Magnetar spindown: The typical spindown timescale for a millisecond magnetar is $10^2-10^4$ s \citep{usov92,zhangmeszaros01a}. If an unsteady magnetized wind is launched, a long GRB with a variable lightcurve can be generated \citep{usov92,thompson94}. In this case, the GRB duration is defined by the spindown timescale. Observationally, both long and short GRBs are followed by X-ray plateaus \citep{zhang06,nousek06,obrien06,lyons10,lvzhang14,lv15}. Some plateaus have the post-plateau decay slope steeper than 3, suggesting that emission has to be ``internal'' and be powered by a long-lived magnetar engine \citep{troja07,rowlinson10}. Also some short GRBs are followed by extended emission lasting for $\sim 100$ s \citep{norris06}. The magnetar spindown power has been widely invoked to interpret these plateaus or extended emission \citep{zhang06,metzger08,lvzhang14,lv15}. So, the GRB prompt emission is most likely powered by the other two mechanisms discussed above.
\end{itemize}

In any case, for magnetar-engine-powered GRBs, the GRB duration no longer necessarily carries the information about the progenitor system. Only when the jet is powered by accretion onto the magnetar could one infer the type of the progenitor system (which we will attempt to do for GRB 211211A, see sub-section \ref{sec:GRB211211A} below).

\subsection{Emitter-defined duration}\label{sec:emitter}

\begin{figure*}[t]
\includegraphics[width=1.0\columnwidth]{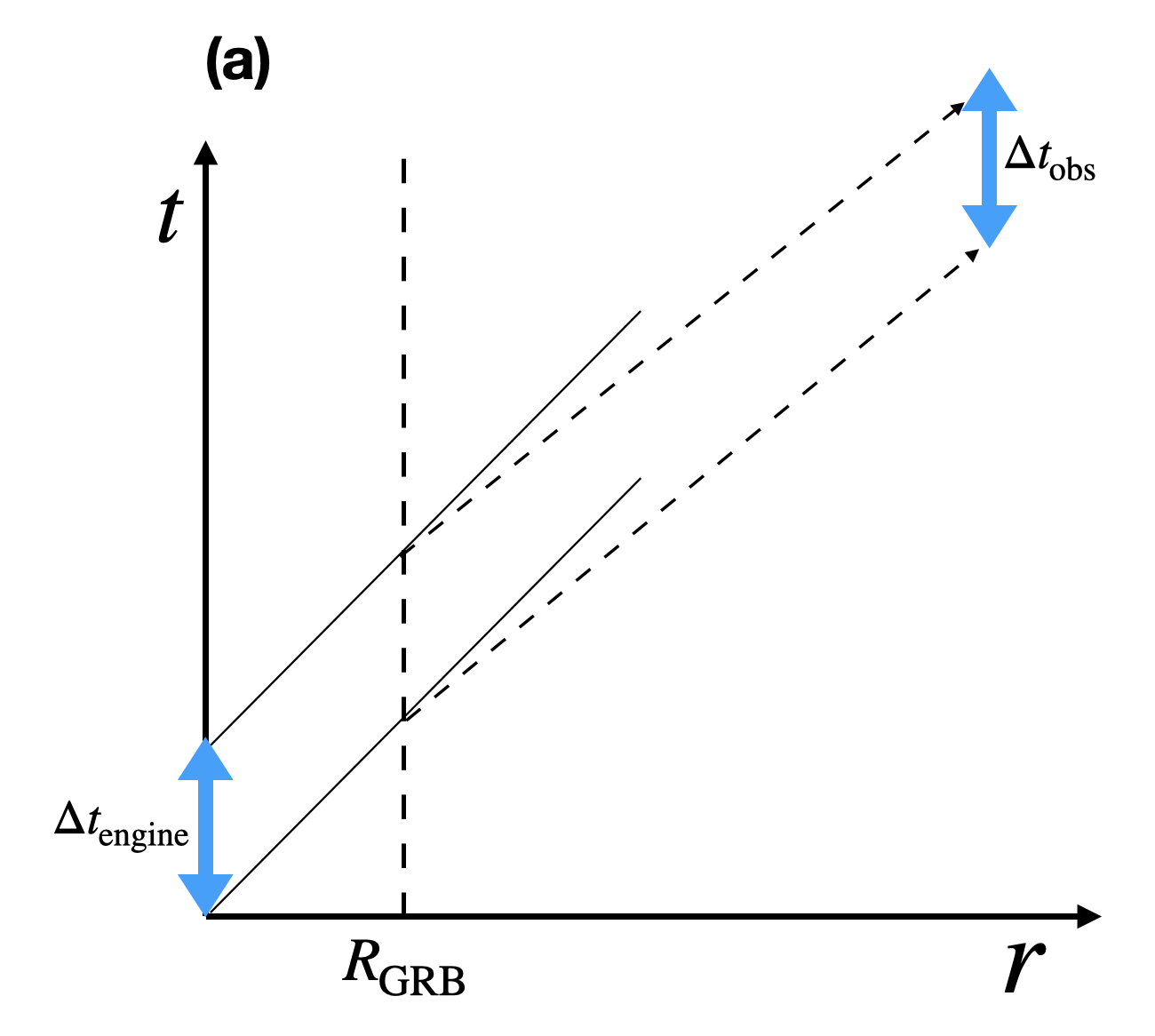}
\includegraphics[width=1.0\columnwidth]{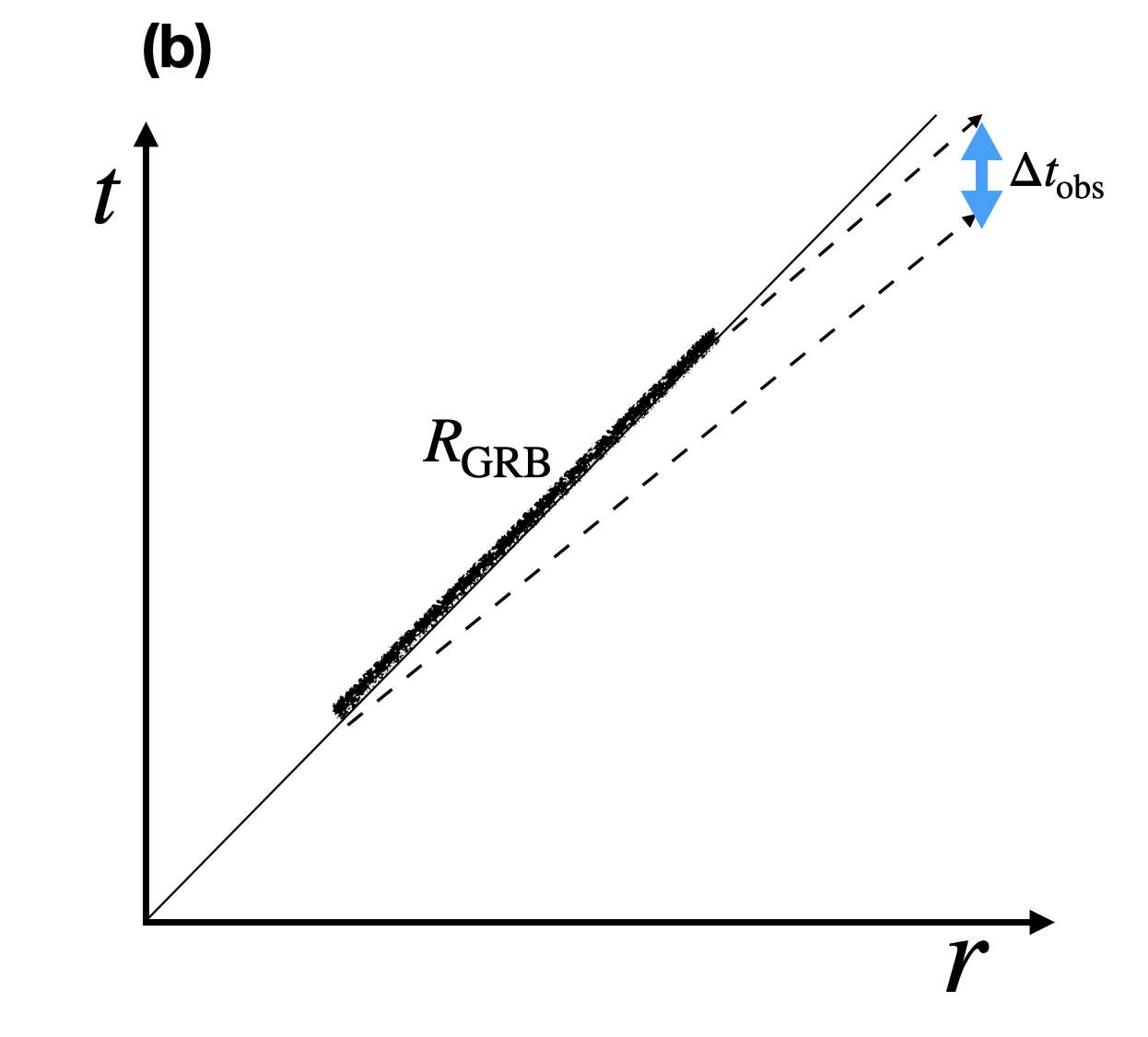}
\caption{The spacetime diagram of GRB emission for two scenarios of duration definition. solid lines denote the sub-luminal motion of the emitter, and dashed arrows denote emitted light rays. (a) The emitters emit instantaneously upon reaching a characteristic emission radius $R_{\rm GRB}$. The observed duration well reflects the duration of the central engine activity. This scenario applies to the internal shock models and the photospheric models. (b) The emitter continues to emit GRB emission over a large range of radii. Even if the central engine duration is instantaneous, the observer could still observe a duration, which is defined by the duration of emission by the emitter. Such a scenario applies to the ICMART model and similar models invoking emission from a single emitter traveling to a large emission radius. }
\label{fig:spacetime}
\end{figure*}

In the previous two subsections, the GRB emitter emission time has a one-to-one correspondance with the central engine time. In other words, each spike in the GRB lightcurve corresponds to an independent emitter launched from the central engine. These emitters release $\gamma$-ray photons when they reach a characteristic radius (e.g. the internal shock radius or the photosphere radius) and shine briefly. The duration of emission of each emitter is shorter than the time separations between different emitters. The spacetime diagram of such a scenario is shown in Figure \ref{fig:spacetime}a, which shows that the observed duration $\Delta t_{\rm obs}$ matches that of the central engine $\Delta t_{\rm engine}$ nicely. Such a scenario applies to the internal shock models \citep{rees94,kobayashi97,daigne98,maxham09} or various forms of the photospheric emission models \citep{meszarosrees00,rees05,peer06,giannios08,beloborodov10,peer11,lazzati11} of GRB prompt emission.

However, there could be another situation. The emitter itself continuously shine in a much longer period of time, so that the observed duration is related to the emission time of the emitter itself rather than the duration of the central engine. As shown in Figure \ref{fig:spacetime}b, a GRB with a certain duration can be observed even for an impulsively injected emitter with $\Delta t_{\rm obs} \gg \Delta t_{\rm engine} \sim 0$. In this case, the lightcurve should have a broad pulse profile, typically have well-defined energy-dependent behaviors, such as the hard-narrow-soft-broad, hard-early-soft-late patterns commonly observed in well-studied pulses in some GRBs \citep{norris02,lu12}. Such a synthetic behavior can be physically understood within the synchrotron radiation model with a decaying magnetic field as the emitter propagates to progressively larger distances \citep{uhm16a,uhm18}. The emission radius of such a model is usually large, typically beyond $10^{15}$ cm, as is expected in the internal collision-induced magnetic reconnection and turbulence (ICMART) model \citep{zhangyan11}. It is worth noting that sometimes rapidly variable spikes are superposed on these broad pulses. This can be readily explained by mini-jet emission in the large emission region, as expected in the ICMART model when a moderately Poynting-flux-dominated outflow with $\sigma > 1$ dissipates energy through reconnection and ejects mini-jets with a Lorentz factor of the order of $\gamma \sim \sqrt{1+\sigma}$. Within this picture, the rapid variability no longer reflects the erratic central engine activity, but is attributed to local energy dissipation processes within a single emitter \citep{zhangyan11,zhangzhang14,shaogao22}. 

Within such a picture, if a GRB only contains one broad pulse with well-behaved energy dependence, then the duration of the GRB could be significantly modified (lengthened) by the emitter. One cannot readily infer the duration of the central engine activity (see the case study of GRB 230307A below in subsection \ref{sec:GRB230307A}). If a GRB contains several broad pulses, even if the duration of each pulse is modified by the emitter, the overall duration, especially the separations between different broad pulses, are still mostly defined by the central engine activity, so the observed duration can be still used to infer the central engine activity time, and even the progenitor type (if the duration is defined by the accretion timescale). 

\subsection{Geometry-defined duration}\label{sec:geometry}

The GRB duration can be also modified by a geometric effect. The effect is significant if the line of sight is outside the relativistically beamed jet cone. For a uniform jet with opening angle $\theta_j$, the observed duration by an off-beam observer at viewing angle $\theta_v>\theta_j$ from the jet axis is related to that of an on-beam observer through
\begin{equation}
    T_{\rm GRB}^{\rm off} = \frac{\cal D_{\rm on}}{\cal D_{\rm off}} T_{\rm GRB}^{\rm on},
\end{equation}
where 
\begin{eqnarray}
    {\cal D}_{\rm on} &\simeq & 2\Gamma, \\
    {\cal D}_{\rm off} & = & \frac{1}{1-\beta\cos(\theta_v-\theta_j)}
\end{eqnarray}
are the on-beam and off-beam Doppler factors, respectively. 

Such a modification may not always happen, however. This is because GRB jets are usually structured, i.e. usually there is a relativistically beamed emitter along the line of sight at large viewing angles, even though both energy per solid angle and bulk Lorentz factor could be much smaller compared with those at the jet core \citep{meszaros98,daigou01,zhangmeszaros02b,rossi02}. Such a structured jet has been inferred from the well-studied GRBs, such as GRB 170817A at the off-axis configuration \citep[e.g.][]{troja18,takahashi20,beniamini22} and the brightest-of-all-time GRB 221009A at the on-axis configuration \citep[e.g.][]{BOAT-structured-jet,BOAT-Gill,zhang24}. In the case of a structured jet, the duration of emission at a particular viewing angle is essentially defined by the emitter along that direction. The GRB duration is therefore defined by the three processes defined in Sections \ref{sec:progenitor}-\ref{sec:emitter}, without significant modifications due to the geometric effect. Only when the jet has a sharp edge will the geometrical modification becomes significant. One example is the very narrow core of the bright jet observed in GRB 221009A that seems to have a sharp edge \citep{BOAT-cao}. An off-beam observer outside the jet cone may detect it as a bright, long-duration X-ray flare, as suggested by \cite{zhang24}.

\section{Case studies}\label{sec:case}

With the above preparations, we are now at the position to study the three peculiar GRBs.

\subsection{GRB 200826A}\label{sec:GRB200826A}

GRB 200826A has $T_{90}$ of 1 s and a large amplitude parameter (defined by the ratio of peak flux $F_p$ over background flux $F_b$) $f \equiv F_p/F_b = 7.53 \pm 1.23$ \citep{zhangbb21}, which is much greater than $f_{\rm eff}$ (the $f$ parameter if the tip-of-iceberg of a long GRB is shorter than 2 s) of long GRBs (typically smaller than 2) \citep{lv14}. This suggests that the GRB duration is indeed quite short. Since a supernova was detected in association with the event \citep{ahumada21,zhangbb21}, this burst is indeed a short-duration Type II GRB. The best interpretation is that ${\rm max}(t_{\rm ff}, t_{\rm acc})$ is only slightly longer than $t_{\rm bo}$ in Equation (\ref{eq:TGRB}). The chance probability for this to happen is small \citep{bromberg11b}, which is consistent with the rareness of similar events. 

\subsection{GRB 211211A}\label{sec:GRB211211A}

GRB 211211A is the first authentic long-duration GRB \citep{yangj22} with a kilonova association \citep{rastinejad22}, so it is clearly a long-duration Type I GRB. Previous long-duration Type I GRBs such as GRB 060614 \citep{gehrels06,zhang07b} still have features similar to known short GRBs with extended emission. They can be still understood within the framework of binary neutron star mergers with a magnetar central engine, with the short-hard spike interpreted as emission powered by accretion and the extended emission interpreted as the signature of magnetar spindown \citep[e.g.][]{metzger08}. 

GRB 211211A pushed this model to the limit. Three features need to be interpreted \citep{yangj22}. First, the prompt emission has a hard main emission (ME) component whose spectral feature is similar to short GRBs, but it lasted for 13 s; second, the prompt emission has a later, softer extended emission (EE) component whose spectral feature is similar to long GRBs, and it lasted for 55 s; finally, the X-ray afterglow shows a plateau feature lasting for $\sim 10^4$ s, which may also require a late engine activity to interpret. A successful progenitor/engine model needs to explain all three features. Below are some models proposed to interpret this event and the issues of these interpretations. 
\begin{itemize}
    \item \cite{zhujp22} interpreted the GRB as an NS-BH merger. The kilonova signature can be well interpreted, but it is hard to use the same BH engine to power three distinct engine episodes (ME, EE, and X-ray plateau) with nearly flat luminosity curves for each component, because a BH engine quickly falls into the $\dot M \propto t^{-5/3}$ delay law due to ejecta fallback. 
    \item \cite{gottlieb23} attempted to interpret GRB 211211A and similar bursts as binary neutron star mergers with a large mass ratio that form a black hole with a massive accretion disk. However, this model suffers from the same criticisms as the NS-BH merger model, because it is hard to have the same BH engine to produce three distinct emission episodes with very different emission behaviors, and the accretion rate is expected to quickly fall off due to fallback. Also this model has some specific predictions (e.g. a $t^{-2}$ decay for extended emission, which has violated observational constraints of many short GRBs with internal plateaus \citep{lv15} and the detailed data of the two long-duration Type I GRBs \citep{yangj22,sun23}. 
    \item The standard binary neutron star merger with a magnetar engine proposed for short GRBs with extended emission also have troubles to interpret GRB 211211A. If the short hard spike (or ME) is powered by accretion, it is unclear why this burst has a much longer ME (13 s) than the majority of short GRBs and short GRBs with extended emission (shorter than 2 s). One possibility is to intepret ME as the magnetic bubble emission due to differential rotation. However, if the $10^4$-s X-ray plateau is a result of magnetar spindown, then two distinct mechanisms are needed to interpret with the ME and the EE that lasted for 55 s. \end{itemize}
By exclusion of other models, \cite{yangj22} proposed an interpretation to the data which can be summarized as follows. The observed duration of GRB 211211A may include the contributions from both progenitor-defined and engine-defined components, with three distinct episodes powered by three distinct physical processes. 1. The ME is powered by accretion onto a magnetar central engine; 2. The EE is powered by magnetic bubble mechanism of the magnetar engine; 3. The X-ray plateau is powered by magnetar spindown. Within this scenario, the accretion episode lasted much longer than most Type I GRBs, which points to a progenitor system with a density falling between neutron stars and massive stars. \cite{yangj22} proposed a white dwarf (WD) - NS merger as the progenitor system for GRB 211211A. Since a magnetar engine needs to be produced upon merger, the WD needs to be near the Chandrasekhar mass limit, so that the merger essentially triggers accretion-induced-collapse (AIC) of the WD. The follow-up procedure is similar to NS-NS mergers, and the kilonova signature can be explained by an engine-fed kilonova \citep{ai22,ai24}. Such a model demands a new type of progenitor not known before, and its confirmation needs to await for the era of space-borne gravitational waves \citep{yin23}. Detailed modeling along this line of reasoning has shown that various prompt emission, afterglow, and kilonova features of the two long-duration Type I GRBs can be well interpreted within the framework of this model \citep{zhong23,zhong24,wangIvy24}\footnote{Note that these models interpret both long-duration Type I GRBs as due to WD-NS mergers. As I argue below in subsection \ref{sec:GRB230307A}, the data of GRB 230307A does not necessarily require a WD-NS merger progenitor.}.

\subsection{GRB 230307A}\label{sec:GRB230307A}

GRB 230307A is another long-duration Type I GRB with kilonova association \citep{sun23,levan24,yang24}. However, its prompt emission is much simpler than GRB 211211A. Even though there are rapidly variable spikes, the envelope of the emission is consistent with one single broad pulse with well-defined energy-dependent pulse widths and spectral lags \citep{yi23}. Such a feature is well consistent with the emitter-defined duration as discussed in Section \ref{sec:emitter}, which means that the intrinsic duration of the central engine activity to power the prompt emission cannot be measured. It is entirely possible that the episode lasted shorter than 2 s (see also \citealt{yi24}). 

The novel observational feature of this burst is that the soft X-ray emission as measured by the Lobster Eye Imager for Astronomy (LEIA) showed a distinct temporal behavior from $\gamma$-rays, clearly suggesting the emergence of a magnetar engine \citep{sun23}. So this burst can be understood within the standard framework of NS-NS mergers with a magnetar engine, even though a progenitor system invoking a near-Chandrasekhar-limit WD-NS merger (the model proposed to interpret GRB 211211A) cannot be excluded. In any case, the models invoking a black hole engine \citep[e.g.][]{zhujp22,gottlieb23} cannot interpret the full data set, especially the global energy-dependent behavior and the robust evidence of the emergence of a magnetar engine, and therefore are disfavored to interpret this GRB. 

\section{Summary}\label{sec:summary}

In this Letter, I revisit various physical processes that shape the duration of a GRB, with the motivation to understand several peculiar GRBs discovered recently. The conclusions can be summarized as follows:
\begin{itemize}
    \item The progenitor-defined GRB duration is only relevant if the GRB jet is powered by accretion. For a black hole engine, one would expect a clean connection between the progenitor systems and and the duration of GRBs.
    \item If the GRB engine is a millisecond magnetar, the duration is defined by the activities of the magnetar (accretion, magnetic bubbles due to differential rotation, unsteady spindown wind), and does not necessarily depend on the progenitor system. Since the clean duration separation between Type I and Type II is muddled, it suggests that at least some GRBs from both types have a magnetar engine. 
    \item GRB duration can be also modified by the emitter, especially if the emitter continuously emits radiation in a large range of emission radii, as expected in models such as ICMART. For a single broad pulse GRB, the observed duration can be only regarded as the upper limit of the duration of the central engine activity. 
    \item GRB duration can be modified geometrically if the line of sight is outside the relativistic beam. Such a situation is rare because of the angular structure of most GRB jets. 
\end{itemize}

The three pecular GRBs can be understood as follows:
\begin{itemize}
    \item The short Type II GRB 200826A is likely a GRB with a relatively short engine activity duration that is slightly longer than the jet breakout time.
    \item The long Type I GRB 230307A can be still understood within the framework of a NS-NS merger with a magnetar engine, with the main emission duration defined by the emitter rather than the progenitor or engine. The magnetar engine launches a Poynting-flux-dominated wind that dissipates energy at a large radius in the form of mini-jet emission, which makes rapid variability on a well-behaved broad pulse (see also \citealt{yi23}). 
    \item The long Type I GRB 211211A presents the most complicated case, with three emission features (ME, EE, and X-ray plateau) demanding possible three processes. Simple models invoking a BH engine or an NS-NS merger with a magnetar engine seem to have trouble to account for the full data set. A near-Chandrasekhar-limit WD-NS merger progenitor is speculated. 
\end{itemize}

\section*{Acknowledgements}

The author acknowledges NASA 80NSSC23M0104 for support and the following colleagues for interesting discussion or helpful comments on various subjects discussed in this Letter: Ore Gottlieb, Shiho Kobayashi, Dong Lai, Andrew Levan, Wenbin Lu, Brian Metzger, Hui Sun, Elenora Troja, Shu-Xu Yi, Bin-Bin Zhang, and especially Zi-Gao Dai as the referee.







\end{document}